\documentclass[aps,prl,twocolumn,groupedaddress,amsmath,amssymb]{revtex4-1}
\usepackage{graphicx,color} 
\usepackage{dcolumn}
\usepackage{bm}
\usepackage{amssymb,amsmath}

\begin{document}
\title{Gate controlled spin-valley locking of resident carriers in WSe$_2$ monolayers}

\author{P. Dey$^1$, Luyi Yang$^1$, C. Robert$^2$, G. Wang$^2$, B. Urbaszek$^2$, X. Marie$^2$, S. A. Crooker$^1$}

\affiliation{$^1$National High Magnetic Field Laboratory, Los Alamos National Lab, Los Alamos, NM 87545, USA}

\affiliation{$^2$Universit\'e de Toulouse, INSA-CNRS-UPS, LPCNO, 135 Av. Rangueil, 31077 Toulouse, France}

\begin{abstract}Using time-resolved Kerr rotation, we measure the spin/valley dynamics of resident electrons and holes in single charge-tunable monolayers of the archetypal transition-metal dichalcogenide (TMD) semiconductor WSe$_2$. In the $n$-type regime, we observe long ($\sim$70~ns) polarization relaxation of electrons that is sensitive to in-plane magnetic fields $B_y$, indicating spin relaxation. In marked contrast, extraordinarily long ($\sim$2$~\mu$s) polarization relaxation of holes is revealed in the $p$-type regime, that is unaffected by $B_y$, directly confirming long-standing expectations of strong spin-valley locking of holes in the valence band of monolayer TMDs. Supported by continuous-wave Kerr spectroscopy and Hanle measurements, these studies provide a unified picture of carrier polarization dynamics in monolayer TMDs, which can guide design principles for future valleytronic devices.
\end{abstract}
\maketitle

Besides their obvious promise for 2D optoelectronics \cite{WangReview, Geim, MakShan},  monolayer transition-metal dichalcogenide (TMD) semiconductors such as MoS$_2$ and WSe$_2$ have also revitalized interests in exploiting both the spin \textit{and valley pseudospin} of electrons and holes for potential applications in (quantum) information processing \cite{Shayegan, graphene, Rycerz, Xiao, XuReview, Mak3}.  This notion of ``valleytronics" arises due to their crystalline asymmetry and strong spin-orbit coupling, which leads to spin-valley locking and valley-specific optical selection rules \cite{XuReview, Xiao}. These rules mandate that the $K$ or $K'$ valleys in momentum space can be selectively populated and probed using polarized light, in contrast with most conventional III-V, II-VI, and group-IV semiconductors. Therefore, information may be readily encoded not only by whether an electron (or hole) has spin ``up" or ``down", but \textit{also} by whether it resides in the $K$ or $K'$ valley -- or, indeed, in some quantum-mechanical superposition thereof.

The intrinsic timescales of carrier spin and valley dynamics in monolayer TMDs are therefore of considerable interest. However, most studies to date \cite{Mai, Singh, Yu, Wang, Zhu, Hao, Schmidt} have focused on photogenerated neutral and charged \textit{excitons}, whose dynamics at low temperatures are inherently limited by their short (3-30~ps) recombination lifetimes \cite{Wang, Robert}. An essential but altogether different question, however, concerns the intrinsic spin/valley lifetimes of the \textit{resident} electrons and holes that exist in $n$-type and $p$-type TMD monolayers. In future valleytronic devices, it is likely the properties of these resident carriers that will determine performance -- analogous to how the scattering timecales and mobility of resident carriers (not excitons) determines the performance of modern-day transistors and interconnects. 

Several recent time-resolved studies point to encouragingly long polarization dynamics of resident carriers in monolayer TMDs.  3-5~ns polarization decays were observed in CVD-grown MoS$_2$ and WS$_2$ monolayers that were unintentionally electron-doped \cite{Luyi, Bushong}, while somewhat longer timescales were observed in unintentionally hole-doped CVD-grown WSe$_2$ \cite{Hsu, Song}. However, a significant shortcoming in all these studies is that the carrier densities were fixed and were due to uncontrolled background impurities.  Therefore, systematic trends with doping density were impossible to identify. It is widely anticipated, however, that spin/valley dynamics in TMD monolayers should depend sensitively on carrier density, \textit{particularly} when tuning between $n$- and $p$-type regimes \cite{Xiao, XuReview}. This is because the huge valence band spin-orbit splitting requires that any $K \leftrightarrow K'$ valley scattering of holes must \textit{also} flip spin (in contrast to the conduction band, where the spin-orbit splitting is much smaller), severely limiting available relaxation pathways \cite{Xiao}.  Time-resolved studies of spin/valley dynamics, in which the carrier concentrations can be systematically tuned between $n$- and $p$-type in a single crystal, are therefore greatly desired, but have not been performed until now. 

Here we experimentally demonstrate and directly validate long-standing predictions of exceptionally robust spin-valley polarization of resident holes in monolayer TMDs. Via time-resolved Kerr rotation studies of \textit{charge-tunable} WSe$_2$ monolayers, we reveal the existence of extraordinarily longlived (2 $\mu$s) spin-valley polarization relaxation in the \textit{p}-type regime. Much shorter polarization dynamics are observed for electrons in the \textit{n}-type regime. Supported by continuous-wave Kerr spectroscopy and Hanle-effect studies, these measurements provide a unified picture of carrier polarization dynamics in the new family of monolayer TMD semiconductors.

Figure 1a depicts the experiment. Exfoliated WSe$_2$ monolayers were transferred onto split Cr/Au gate electrodes patterned on SiO$_2$/Si substrates \cite{SM}. Compared to typical CVD-grown WSe$_2$, WS$_2$, or MoS$_2$, exfoliated WSe$_2$ exhibits fewer defects and much cleaner optical spectra.  Of particular importance, and in contrast to earlier studies of resident carrier dynamics \cite{Luyi, Bushong, Hsu, Song}, the neutral ($X^0$) and charged exciton ($X^\pm$) transitions are spectrally resolved and can be probed separately. A gate voltage $V_G$ tunes between $n$-type and $p$-type regimes, as confirmed in Figs. 1b,c by low-temperature reflection spectra.  When nominally undoped ($V_G$=0), the spectra exhibit a single resonance at $\sim$712~nm corresponding to neutral ``A" excitons, $X^0$. However when $V_G$$<$0 and the conduction bands fill with resident electrons, the $X^0$ resonance weakens and the negatively-charged exciton ($X^-$) resonance develops at $\sim$725~nm.  The $X^-$ is a three-particle complex consisting of a photogenerated electron-hole pair bound to an additional resident electron. Similarly, when $V_G$$>$0 and the valence bands fill with resident holes, the spectra evolve into the positively-charged exciton $X^+$ (an electron-hole pair bound to an additional resident hole) at $\sim$720~nm. This behavior agrees very well with past studies of gated WSe$_2$ monolayers \cite{Jones}.

The gated samples were studied using both continuous-wave and time-resolved Kerr rotation (CWKR, TRKR); Fig. 1a depicts the latter \cite{SM}. Right- or left-circularly polarized (RCP/LCP) pump pulses photoexcite excess spin-up or spin-down electrons and holes in the $K$ or $K'$ valley, respectively. At cryogenic temperatures, these photocarriers form excitons and trions that quickly scatter and recombine on short ($<$30~ps) timescales, as determined by time-resolved PL \cite{Wang, Robert}. In doing so they perturb the resident carriers away from thermal equilibrium, thereby transferring to them a non-zero spin/valley polarization. The mechanisms underpinning polarization transfer likely include exciton correlations and scattering while the photo-generated minority species are present \cite{Mai, Singh, Yu, Schmidt}, and any non-radiative recombination of the minority species with resident carriers having opposing spin and/or valley. Related processes in conventional III-V and II-VI semiconductors are well-known to transfer spin polarizations to resident carriers in bulk, 2D, and 0D systems \cite{Dzhioev, Furis, YakovlevChapter, Cundiff, Atature, Greilich}. Regardless of the generation mechanism (which is not our focus), once the photoexcited carriers have recombined on short timescales, the spin/valley polarization of the resident carriers is \textit{out of} equilibrium and will relax according to its intrinsic and much longer timescales. Our principal goal is to measure these fundamental intrinsic timescales in both electron- and hole-doped regimes, in the same TMD crystal.

The spin/valley polarization is monitored via the Kerr rotation $\theta_K$ imparted on time-delayed probe pulses from a tunable Ti:sapphire laser. An essential new aspect of this experiment is a fast electronic delay generator synchronized to an acousto-optic pulse-picker \cite{SM}. In contrast to prior TRKR studies \cite{Luyi, Bushong, Hsu, Song}, this combination enables pump-probe delays $\Delta t$ up to \textit{microseconds}, which greatly exceeds that of conventional optical delay lines, allowing direct access to the extremely long relaxation timescales that, we find, exist in monolayer WSe$_2$. 

\begin{figure} [tbp]
\includegraphics [width=0.48\textwidth] {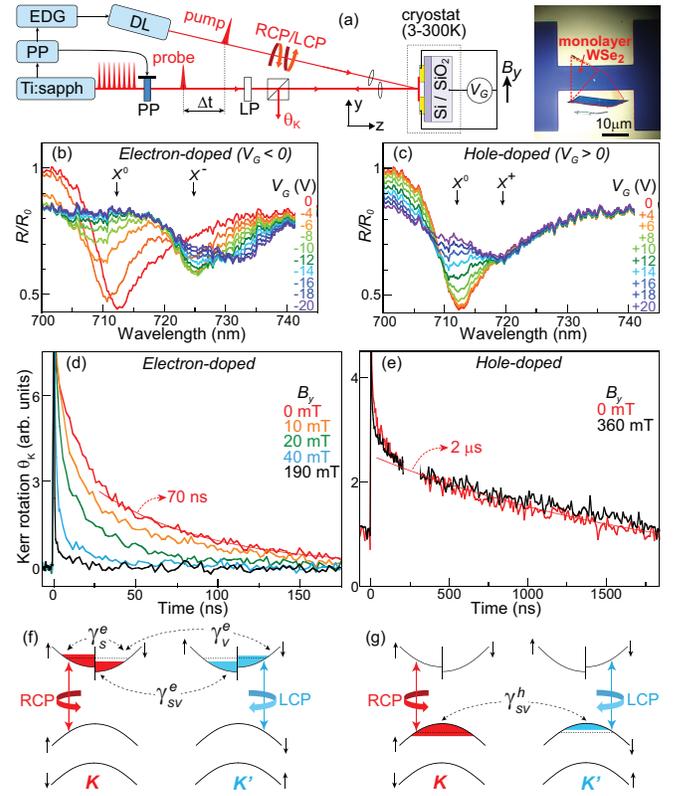}
\caption{(a) TRKR experiment on gated monolayer WSe$_2$. $V_G$ tunes the resident carrier density between electron- and hole-doped regimes. External coils provide in-plane magnetic fields $B_y$. RCP or LCP pump pulses from a 645~nm diode laser (DL) imprint a spin/valley polarization on the resident carriers, which is detected via the Kerr rotation $\theta_K$ imparted on probe pulses from a Ti:S laser. A pulse-picker (PP) and electronic delay generator (EDG) allow access to $\mu$s pump-probe delays $\Delta t$. (b) Normalized reflection spectra $R/R_0$ versus $V_G$ at 5~K; the oscillator strength evolves from neutral exciton $X^0$ to negatively-charged exciton $X^-$ with increasing electron density ($V_G<0$). (c) Similar evolution from $X^0$ to positively-charged exciton $X^+$ with increasing hole density ($V_G>0$). (d) TRKR in the electron-doped regime at 5K, with the probe tuned near $X^-$ (727 nm): The smooth red line denotes a 70~ns exponential decay. The decays shorten dramatically in small $B_y$. (e) TRKR at 5K in the hole-doped regime at the $X^+$ transition: Extraordinarily long polarization decays are revealed ($\sim$2~$\mu$s) that are \textit{independent} of $B_y$, confirming strong spin-valley locking in the valence band. (For technical reasons, $\Delta t$ between 205-310~ns are not accessible). The red line denotes a 2~$\mu$s exponential decay. (f,g) The diagrams depict the simplest WSe$_2$ band structure in the electron and hole-doped regimes, along with available scattering pathways.}
\label{Fig1}
\end{figure}

Figures 1d,e show the central result. In the electron-doped regime, with the probe laser tuned to the $X^-$ resonance, TRKR reveals surprisingly long polarization decays of $\sim$70~ns at 5~K. Crucially, these decays shorten dramatically in weak applied in-plane magnetic fields $B_y$$<$100~mT, strongly suggesting electron spin relaxation as the origin of this decay (discussed below). Most remarkably, however, when the same WSe$_2$ monolayer is populated with holes, TRKR studies with the probe at $X^+$ reveal extraordinarily long-lived polarization decays with a slow component of $\sim$2~$\mu$s. However, these slow decays are \emph{not} affected by $B_y$, consistent with strong spin-valley locking in the valence band.  

To help understand these contrasting behaviors, Figs. 1f,g depict the simplest WSe$_2$ bandstructure and possible relaxation pathways. Three pathways for resident electrons are identified: i) spin relaxation within a valley, given by rate $\gamma^e_s$, ii) spin-\textit{conserving} intervalley scattering ($\gamma^e_v$), and iii) spin-\textit{flip} intervalley scattering ($\gamma^e_{sv}$). The latter rate requires \textit{both} spin and valley scattering, and is therefore likely small. The measured 70 ns relaxation time is influenced by the electron density and how the chemical potential $\mu$ compares to the conduction band spin-orbit splitting, $\Delta_c$ ($\sim$25~meV in WSe$_2$ \cite{Heinz}). In contrast, for resident holes the giant valence band spin-orbit splitting ($\Delta_v$$\sim$450~meV) ensures that $\mu \ll \Delta_v$. Thus polarized holes can only relax by simultaneously scattering both spin \textit{and} valley degrees of freedom. The corresponding rate, $\gamma^h_{sv}$, is therefore expected to be quite small \cite{Xiao, XuReview}. 

The markedly different dependence on $B_y$ provides important information. In the $n$-type regime, the sensitivity to small $B_y$, and absence of any oscillatory TRKR signal, are consistent with a spin depolarization mechanism recently proposed for electron-doped MoS$_2$ \cite{Luyi}, summarized here as follows: $\Delta_c$ is `seen' by resident electrons as a \textit{valley-dependent} effective spin-orbit field $B_{so}$ oriented normal to the 2D plane (parallel to $ \pm \hat{z}$, depending on whether the electron resides in $K$ or $K'$). Being large (10s of tesla), $B_{so}$ should stabilize an electron's spin along $\pm \hat{z}$, such that $B_y$ has little effect. However, if spin-conserving $K \leftrightarrow K'$ scattering ($\gamma^e_v$) is fast (as believed for electrons in TMDs \cite{Kis, Mai}), then $B_{so}$ fluctuates rapidly between $\pm \hat{z}$.  Therefore, the net field $B_y \hat{y} \pm B_{so} \hat{z}$ is both fluctuating \textit{and} slightly canted, which will quickly depolarize electron spin, as experimentally observed. Within this model, the slow decay measured when $B_y$=0 is $1/ \gamma^e_s$, the intravalley spin relaxation time. 

In contrast, in the $p$-type regime $\Delta_v$ is huge, and spin-conserving intervalley scattering is suppressed as discussed above. $B_{so}$ `seen' by holes is gigantic ($>$10$^3$~T) and does not fluctuate.  Hole spins are pinned along $+ \hat{z}$ or $- \hat{z}$ and $B_y$ should have little influence, as observed. The only available relaxation path is spin-flip valley scattering ($\gamma^h_{sv}$), measured to be $\sim$2~$\mu$s at 5~K. These results experimentally confirm the widely-believed theoretical prediction \cite{Xiao, XuReview} that spin-valley locking in the valence band leads to extremely robust spin-valley polarization of holes. 

The remarkably stable hole polarizations in our WSe$_2$ monolayers are strongly supported by a recent report of microsecond hole polarizations of indirect excitons in WSe$_2$/MoS$_2$ \textit{bilayers} \cite{FengWang}.  Here, rapid electron-hole spatial separation following neutral exciton generation leads to long-lived indirect excitons, in which the electron and hole are very weakly bound and therefore depolarize essentially as independent particles, approximating the case of resident carriers.  Our experiments on gated WSe$_2$ monolayers directly confirm that resident holes indeed have intrinsically long polarization lifetimes due to spin-valley locking which is not limited by population decay. This suggests monolayer WSe$_2$ is an excellent building block for hole spin/valley storage in more complex van der Waals devices \cite{Geim}.

Moreover, the very different dynamics in $n$- and $p$-type regimes argues against these long decays being due to optically-forbidden (``dark") neutral excitons \cite{Plechinger}, which are believed to exist in WSe$_2$ but which likely decay on shorter timescales \cite{Heinz}. We note, however, that putative \textit{charged} dark excitons could in principle  play a role in these gate-dependent studies.

\begin{figure} [tbp]
\includegraphics [width=0.48\textwidth] {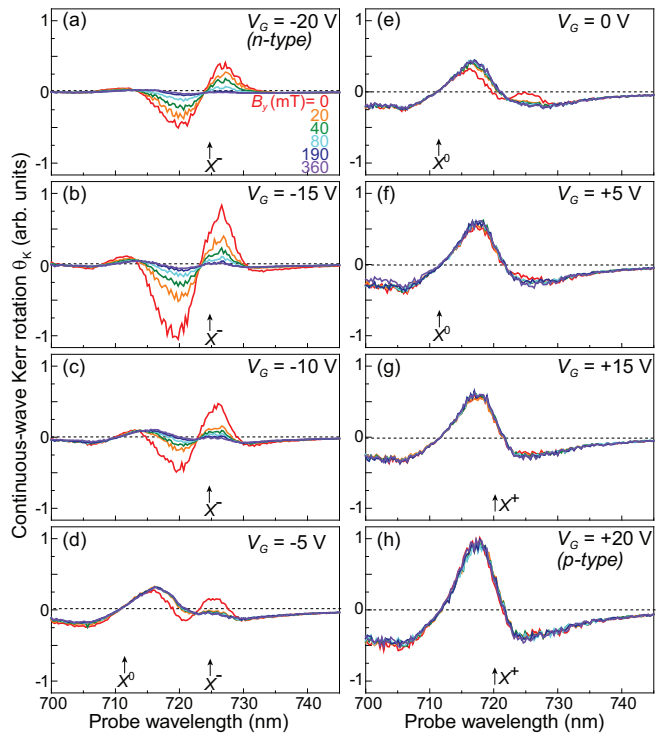}
\caption{(a-h) CWKR spectra at 5~K as the resident carrier density is tuned from the electron-doped ($V_G=-20$~V) to nominally undoped ($V_G=0$~V) to hole-doped ($V_G=+20$~V) regime.  Within each panel, $B_y$ is varied from 0~mT (red trace) to 360~mT (violet trace). The WSe$_2$ is weakly pumped by a circularly-polarized 632.8~nm CW laser, and the induced Kerr rotation is detected by a narrowband CW Ti:sapphire ring laser that is scanned from 700-750~nm. Positions of the neutral and charged exciton transitions from reflectivity studies (see Figs. 1b,c) are indicated.}
\label{Fig2}
\end{figure}

Our TRKR results are supported by detailed CWKR spectroscopy, wherein a weak CW pump generates a \textit{steady-state} non-equilibrium carrier polarization while a narrowband probe, scanned across the neutral and charged exciton transitions, detects $\theta_K$ \cite{SM}. Figure 2 shows CWKR spectra at various $V_G$ spanning $n$- to $p$-type doping regimes. At each $V_G$, spectra are measured at different $B_y$ (red to violet curves).  When unambiguously electron-doped ($V_G$$\ll$0; Figs. 2a-c), the CWKR spectra exhibit a sizable Kerr resonance centered at $\sim$725~nm, which corresponds to the $X^-$ transition. The antisymmetric Kerr lineshape is typical, and reveals a large steady-state polarization. Consistent with TRKR studies, small $B_y$ suppresses $\theta_K$. As $V_G \rightarrow 0$ and the electrons deplete, a smaller Kerr resonance develops at the $X^0$ transition ($\sim$710~nm); however, it is largely unaffected by $B_y$, as expected \cite{Sallen}. 

Crucially, these data show that the resident carrier polarization appears to manifest itself primarily via the \textit{charged} exciton transition. Analogous to the well-studied situation in III-V and II-VI semiconductors \cite{YakovlevChapter, Cundiff, Atature, Greilich}, this is likely because the probability for exciting a polarized $X^-$ \textit{necessarily} depends on the availability of appropriately-polarized resident electrons (or resident holes, for $X^+$ formation).  Consider the simplest case of singlet $X^-$, wherein the photogenerated electron must have spin orientation opposite to the resident electron's. In the limit where the resident electrons are entirely polarized spin-up, the absorption of (say) RCP light at $X^-$ will be entirely suppressed, while the absorption for LCP light remains large. This circular dichroism generates a large $\theta_K$. Kerr rotation at the charged exciton transitions is therefore a sensitive probe of the resident carrier polarization. In contrast, $X^0$ formation does not depend explicitly on the polarization of a third (resident) particle.  This further highlights the importance of using exfoliated WSe$_2$: resolving the very different trends at $X^0$ and $X^-$ provides essential insight into the underlying mechanisms. In many recent TRKR studies of CVD-grown MoS$_2$, WS$_2$, and WSe$_2$ \cite{Luyi, Bushong, Hsu, Song}, a clear distinction between neutral and charged excitons was not possible. 

Returning to Fig. 2, when the WSe$_2$ is $p$-type ($V_G$$\gg$0) the CWKR spectrum distorts, grows, and develops a sharp zero-crossing at $\sim$720~nm, which corresponds to the $X^+$ transition. Consistent with the TRKR data, these large CW Kerr signals reflect a buildup of steady-state hole polarization, and -- importantly -- are \textit{not} influenced by $B_y$ due to spin-valley locking. 

Finally, we perform Hanle effect measurements (\textit{i.e.}, carrier depolarization by $B_y$), which have historically played a central role in semiconductor spintronics \cite{Dzhioev, Furis} to determine spin lifetimes, nuclear fields, and spin-orbit effects. Figure 3 shows $\theta_K$ versus $B_y$ as the WSe$_2$ monolayer is tuned from $n$- to $p$-type. When probing at 727~nm -- where both $X^-$ and $X^+$ give appreciable $\theta_K$ -- the signals in the electron-doped regime exhibit narrow peaks indicating that small $B_y$ dephases the electron polarization (as known from TRKR studies). Interestingly, however, the peak widths are narrowest (18~mT) when lightly electron-doped, but increase to $>$60~mT at higher electron densities.  As discussed in \cite{Luyi}, the Hanle width in monolayer TMDs does not simply reflect the spin lifetime (as in conventional semiconductors), but rather depends on the interplay between spin relaxation $\gamma^e_s$, intervalley scattering $\gamma^e_v$, and $B_{so}$. The observed trend is consistent with increased spin-orbit splitting `seen' by electrons as the chemical potential rises, because the spin-up and spin-down bands have different curvatures. Conversely, in the hole-doped regime the traces are flat because $B_y$ does not affect the hole polarization due to spin-valley locking. Further, Fig. 3b shows Hanle studies at the $X^0$ resonance (706~nm).  Here, $\theta_K$ is essentially unaffected by $B_y$ for \textit{all} $V_G$, again indicating that $X^0$ is uninfluenced by the resident carrier polarization. 

\begin{figure} [tbp]
\includegraphics [width=0.35\textwidth] {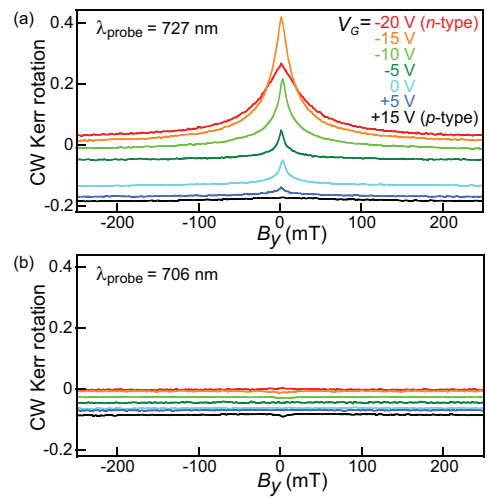}
\caption{(a) Hanle-Kerr effect as the WSe$_2$ monolayer is tuned from $n$-type to $p$-type regime. The WSe$_2$ is weakly pumped by a circularly-polarized 632.8~nm CW laser. $B_y$ is continuously varied while the induced Kerr rotation is measured by a tunable CW probe laser. Here the probe is fixed at 727~nm, which is sensitive to both $X^-$ and $X^+$ transitions. The width of the Hanle curves increases with increasing electron density in the $n$-type regime. In the $p$-type regime the Hanle curves show no dependence on $B_y$. (b) Same, but the probe laser is fixed at 706~nm, which is sensitive to the neutral $X^0$ exciton transition. No dependence on $B_y$ observed.}
\label{Fig3}
\end{figure}

Taken all together, these time-resolved and steady-state Kerr studies of single gated TMD monolayers significantly advance a unified picture of spin and valley dynamics of resident carriers in atomically-thin semiconductors. Controlling strong spin-valley locking and ultralong polarization storage lifetimes by tuning between $n$- and $p$-type doping will likely be significant for future TMD-based valleytronics using van der Waals devices. 

P.D., L.Y., and S.A.C. gratefully acknowledge support from the Los Alamos LDRD program and the NHMFL, which is supported by NSF DMR-1157490.  C.R. and X.M. thank ANR MoS$_2$ValleyControl, G.W. and B.U. thank ERC Grant No. 306719; X.M. also acknowledges the Institut Universitaire de France. We also thank M. Pierre, W. Escoffier and B. Lassagne for device fabrication help, and S. Tongay for the growth of bulk WSe$_2$.


\begin{references}

\bibitem{WangReview} Q. H. Wang, K. Kalantar-Zadeh, A. Kis, J. N. Coleman, and M. S.  Strano, Nat. Nanotechnol. \textbf{7}, 699 (2012).

\bibitem{Geim} A. K. Geim and I. V. Grigorieva, Nature \textbf{499}, 419 (2013).

\bibitem{MakShan} K. F. Mak and J. Shan, Nat. Photon. \textbf{10}, 216 (2016).

\bibitem{Shayegan} O. Gunawan, Y. P. Shkolnikov, K. Vakili, T. Gokmen, E. P. De Poortere, and M. Shayegan, Phys. Rev. Lett. \textbf{97}, 186404 (2006).

\bibitem{graphene} D. Xiao, W. Yao, and Q. Niu, Phys. Rev. Lett. \textbf{99}, 236809 (2007).

\bibitem{Rycerz} A. Rycerz, J. Tworzyd\l o, and C. W. J. Beenakker, Nat. Phys. \textbf{3}, 172 (2007).

\bibitem{Xiao} D. Xiao, G. B. Liu, W. X. Feng, X. Xu, and W. Yao, Phys. Rev. Lett. \textbf{108}, 196802 (2012).

\bibitem{XuReview} X. Xu, W. Yao , D. Xiao, and T. F. Heinz, Nat. Phys. \textbf{10}, 343 (2014).

\bibitem{Mak3} K. F. Mak, K. L. McGill, J. Park, and P. L. McEuen, Science \textbf{344}, 1489 (2014).

\bibitem{Mai} C. Mai, A. Barrette, Y. Yu, Y. G. Semenov, K. W. Kim, L. Cao, and  K. Gundogdu, Nano Lett. \textbf{14}, 202 (2014).

\bibitem{Singh} A. Singh, G. Moody, S. Wu, Y. Wu, N. J. Ghimire, J. Yan, D. G. Mandrus, X. Xu, and X. Li, Phys. Rev. Lett. \textbf{112}, 216804 (2014).

\bibitem{Yu}T. Yu and M. W. Wu, Phys. Rev. B \textbf{89}, 205303 (2014).

\bibitem{Wang} G. Wang, L. Bouet, D. Lagarde, M. Vidal, A. Balocchi, T. Amand, X. Marie, and B. Urbaszek, Phys. Rev. B \textbf{90}, 075413 (2014).

\bibitem{Zhu} C. R. Zhu, K. Zhang, M. Glazov, B. Urbaszek, T. Amand, Z. W. Ji, B. L. Liu, and X. Marie, Phys. Rev. B \textbf{90}, 161302(R) (2014).

\bibitem{Hao} K. Hao, G. Moody,	F. Wu,C. K. Dass,	L. Xu,	C.-H. Chen,	L. Sun,	M.-Y. Li,	L.-J. Li,	A. H. MacDonald, and X. Li, Nat. Phys. \textbf{12}, 677 (2016). 

\bibitem{Schmidt} R. Schmidt, G. Berghauser, R. Schneider, M. Selig, P. Tonndorf, E. Malic, A. Knorr, S. M. de Vasconcellos, and R. Bratschitsch, Nano Lett. \textbf{16}, 2945 (2016). 

\bibitem{Robert} C. Robert, D. Lagarde, F. Cadiz, G. Wang, B. Lassagne, T. Amand, A. Balocchi, P. Renucci, S. Tongay, B. Urbaszek, and X. Marie, Phys. Rev. B \textbf{93}, 205423 (2016).

\bibitem{Luyi} L. Yang, N. A. Sinitsyn, W. Chen, J. Yuan, J. Zhang, J. Lou, and S. A. Crooker, Nat. Phys. \textbf{11}, 830 (2015).

\bibitem{Bushong} E. J. Bushong, Y. Luo, K. M. McCreary, M. J. Newburger, S. Singh, B. T. Jonker, and R. K. Kawakami, Preprint at arXiv:1602.03568 (2016).

\bibitem{Hsu} W.-T. Hsu, Y.-L. Chen, C.-H. Chen, P.-S. Liu, T.-H. Hou, L. J. Li, and W.-H. Chang, Nat. Commun. \textbf{6}:8963 (2015).

\bibitem{Song} X. Song, S. Xie, K. Kang, J. Park, and V. Sih, Nano Lett. \textbf{16}, 5010 (2016).

\bibitem{SM}See Supplemental Material at XXXX for experimental details and supporting data. 

\bibitem{Jones} A. M. Jones, H. Yu,	N. J. Ghimire, S. Wu,	G. Aivazian, J. S. Ross,	B. Zhao, J. Yan, D. G. Mandrus,	D. Xiao, W. Yao, and X. Xu, Nat. Nanotech. \textbf{8}, 634 (2013).

\bibitem{Dzhioev} R. I. Dzhioev, K. V. Kavokin, V. L. Korenev, M. V. Lazarev, B. Ya. Meltser, M. N. Stepanova, B. P. Zakharchenya, D. Gammon, and D. S. Katzer, Phys. Rev. B \textbf{66}, 245204 (2002). 

\bibitem{Furis} M. Furis, D. L. Smith, S. Kos, E. S. Garlid, K. S. M. Reddy, C. J. Palmstrom, P. A. Crowell, and S A Crooker, New J. Phys. \textbf{9}, 347 (2009).

\bibitem{YakovlevChapter} D. R. Yakovlev and M. Bayer, Chapter 6, Spin Physics in Semiconductors, M. I. Dyakonov, editor (Springer, Berlin 2008).

\bibitem{Cundiff} Z. Chen, S. G. Carter, R. Bratschitsch, and S. T. Cundiff, Physica E \textbf{42}, 1803 (2010).

\bibitem{Atature} M. Atat\"{u}re, J. Dreiser, A. Badolato, A. H\"{o}gele, K. Karrai, and A. Imamoglu, Science \textbf{312}, 551 (2006).
    
\bibitem{Greilich} A. Greilich, D. R. Yakovlev, A. Shabaev, Al. L. Efros, R. Oulton, V. Stavarache, D. Reuter, A. Wieck, and M. Bayer, Science \textbf{313}, 341 (2006).

\bibitem{Heinz} X. X. Zhang, T. Cao, Z. Lu, Y.-C. Lin, F. Zhang, Y. Wang, Z. Li, J. C. Hone, J. A. Robinson, D. Smirnov, S. G. Louie, and T. F. Heinz, preprint at arXiv:1612.03558


\bibitem{Kis} B. Radisavljevic and A. Kis, Nat. Mater. \textbf{12}, 815 (2013).

\bibitem{FengWang} J. Kim, C. Jin, B. Chen, H. Cai, T. Zhao, P. Lee, S. Kahn, K. Watanabe, T. Taniguchi, S. Tongay, M. F. Crommie, and F. Wang, preprint at arXiv:1612.05359

\bibitem{Plechinger} G. Plechinger, P. Nagler, A. Arora, R. Schmidt, A. Chernikov, A. Granados del Aguila, P.C.M. Christianen, R. Bratschitsch, C. Sch\"{u}ller, and T. Korn, Nat. Commun. \textbf{7}:12715 (2016).

\bibitem{Sallen} G. Sallen, L. Bouet, X. Marie, G. Wang, C. R. Zhu, W. P. Han, Y. Lu, P. H. Tan, T. Amand, B. L. Liu, and B. Urbaszek, Phys. Rev. B \textbf{86}, 081301(R) (2012).

\end{references}
\end{document}